\documentclass[usenatbib]{aa} 

\usepackage{graphicx} 
\usepackage{amssymb} 
\usepackage{epsfig} 
\usepackage{natbib} 
\bibpunct{(}{)}{;}{a}{}{,} 
 
\def\pl{{\sc pl}} 
\def\compps{{\sc compps}} 
\def\kte{kT_{\rm e}} 
\def\taut{\tau_{\rm T}} 
\def\chiq{$\chi^2$}

\def\nh{ {$N_{\rm H}$}}

\def\thesource{IGR~J00291+5934} 
\def\nudot{\dot\nu}

\def\be{\begin{equation}} 
\def\ee{\end{equation}} 
\def\beq{\begin{eqnarray}} 
\def\eeq{\end{eqnarray}} 
 
\newcommand{\gtap}{\mathrel{\hbox{\rlap{\lower.55ex \hbox {$\sim$}} 
                   \kern-.3em \raise.4ex \hbox{$>$}}}} 
\newcommand{\ltap}{\mathrel{\hbox{\rlap{\lower.55ex \hbox {$\sim$}} 
                   \kern-.3em \raise.4ex \hbox{$<$}}}}

\begin{document} 
%\received{Accepted 18 August, 2005}  
\title{ {\it INTEGRAL} and  {\it RXTE} observations of accreting millisecond pulsar IGR 
  J00291+5934 in outburst} 
\author{M. Falanga\inst{1,2}\fnmsep\thanks{\email{mfalanga@cea.fr}}, 
L. Kuiper\inst{3}, 
J. Poutanen\inst{4}, 
E. W. Bonning\inst{5}, 
W. Hermsen\inst{3,6}, 
T. Di Salvo\inst{7}, 
P. Goldoni\inst{1,2}, 
A. Goldwurm\inst{1,2}, 
S. E. Shaw\inst{8,9}, 
L. Stella\inst{10} 
} 
 
\offprints{M. Falanga} 
\titlerunning{INTEGRAL observation of IGR J00291+5934} 
\authorrunning{M. Falanga et al.} 
 
\institute{CEA Saclay, DSM/DAPNIA/Service d'Astrophysique (CNRS FRE 
  2591), F-91191, Gif sur Yvette, France 
\and Unit\'e mixte de recherche Astroparticule et 
Cosmologie, 11 place Berthelot, 75005 Paris, France 
\and SRON Netherlands Institute for Space Research, Sorbonnelaan 2, 
3584 CA Utrecht, The Netherlands 
\and  Astronomy Division, P.O.Box 3000, FIN-90014 University of 
  Oulu, Finland 
\and Laboratoire de l'Univers et de ses Th'{e}ories, Observatoire de 
Paris, F-92195 Meudon Cedex, France 
\and Astronomical Institute ``Anton Pannekoek'', University of 
Amsterdam, Kruislaan 403, NL-1098 SJ Amsterdam, The Netherlands 
\and Dipartimento di Scienze Fisiche ed Astronomiche, Universit\`a 
   di Palermo, via Archirafi 36, 90123 Palermo, Italy 
\and School of Physics and Astronomy, University of 
Southampton, SO17 1BJ, UK 
\and INTEGRAL Science Data Centre, CH-1290 Versoix, Switzerland 
\and Osservatorio Astronomico di Roma, via Frascati 33, 00040 
  Monteporzio Catone, Italy 
} 
 
\abstract{ 
Simultaneous observations of the 
 accretion-powered millisecond pulsar \thesource\ 
by {\it International Gamma-Ray Astrophysics Laboratory} and 
{\it Rossi X-ray Timing Explorer} during the 2004 December outburst 
 are analysed. 
The average spectrum is well described by thermal Comptonization 
 with an electron temperature of 50 keV and Thomson optical depth 
$\taut\sim1$ in a slab geometry. The spectral shape is almost constant 
during the outburst. 
We detect a spin-up of the pulsar with $\nudot=8.4\times 
10^{-13}\mbox{Hz s}^{-1}$. 
 The ISGRI data reveal the pulsation of X-rays at a period of 1.67 
 milliseconds up to $\sim150$ keV. 
 The pulsed fraction is shown to increase from 6 per cent 
 at 6 keV to 12--20 per cent at 100 keV. 
 This is naturally   explained by the action of the Doppler effect  the 
 exponentially cutoff Comptonization spectrum from the hot spot. 
 The nearly sinusoidal pulses show soft lags with 
complex energy dependence, increasing up to 7 keV, then decreasing to 15 keV, 
 and  seemingly saturating at higher energies.
 } 
\maketitle 
 
\section{Introduction} 
\label{sec:intro} 
 
\thesource\ was discovered in the galactic disk 
by the {\it International Gamma-Ray Astrophysics Laboratory} ({\em INTEGRAL}) 
\citep{e04} during a routine Galactic Plane Scan of the Cas A 
region on  December 2, 2004  during an  outburst.   Follow-up 
observations on  December 3, 2004  with the 
{\it Rossi X-ray Timing Explorer} ({\em RXTE}) classified the source 
as an accreting X-ray millisecond pulsar (MSP), the sixth known 
source of this class \citep{mss04}. \thesource\ has  a spin period of 
1.67 ms, making it the fastest known X-ray MSP to 
date, and an orbital period of 2.5 hours \citep{mgc04}. The other known 
X-ray accreting  millisecond pulsars' spin frequencies lie between 180 
and 435 Hz (see review by \citealt*{w05}). Their orbital periods fall 
into two distinct ranges - either around 40 minutes or 2--4.5 hours. 
 
A candidate optical counterpart was observed with diminishing intensity 
 from magnitude $R\sim17-21$, consistent with the X-ray 
flux decay rate of  \thesource\ during the latter days of the outburst 
 \citep{fg04,bsg04,sbb04}. 
It is the second X-ray MSP (after SAX J1808.4-3658) for which the
 optical flux decay could be observed.
Spectroscopy of the optical 
source led to the detection of broad emission lines of HeII and H$\alpha$ 
 \citep{rjs04,ffc04}. Radio observations 
also detected a source consistent with the position of \thesource\ 
 \citep{po04,fd04,rd04}. The source has been observed in quiescence 
 with {\em Chandra} by \citet{jonker05}, and a first {\em INTEGRAL} spectrum 
was reported  by \citet{shaw05}. 
 
The hard X-ray emission of \thesource\ was significantly detected  by 
{\em INTEGRAL} during the entire  fifteen-day observation from  outburst 
to quiescence.  In this paper we analyze the X-ray spectrum during the 
outburst. Furthermore, we perform timing analysis on the {\em INTEGRAL} 
high energy data as well as on data from a {\it RXTE} follow-up 
observation, in order to study the characteristics of the pulse
profile (shape,  time lags, pulsed fraction) at energies from 2 to 150 keV.

\section{Observations and data}

\subsection{INTEGRAL} 
\label{sec:integral} 
 
The  dataset was obtained with {\em INTEGRAL}  \citep{w03} from 
December 2 to December 16 2004, i.e. from satellite 
revolution 261 to 265 (see  Table \ref{tab:obs}). These revolutions 
include a Target of Opportunity (ToO) observation and part of a Cas 
A/Tycho AO2 observation. 
 
\begin{table*}[ht] 
\begin{center} 
\caption{\label{tab:obs} Log of {\em INTEGRAL} observations analysed in this 
  paper. Start and end times are in days of December 
  2004 (from MJD 53340). Exposures are in seconds and count rates are 
  in counts per second, for ISGRI in the 20--100 keV and for JEM-X in the 
  5--20 keV energy bands.} 
\begin{tabular}{rrrrrrrl} 
\hline 
 & &   \multicolumn{2}{c}{ISGRI} & \multicolumn {2}{c}{JEM-X}  &  &\\ 
Start &  End & Exposure & Count rate & Exposure & Count rate & Rev. & 
      Obs.\\ 
\hline 
%1.281  & 1.362  & 4534   & 9.1$\pm$0.8 & 1512  & 2.5$\pm$0.4 & 261 & GPS\\ 
3.671  & 3.840  & 8449   & 8.3$\pm$0.5 & 3023  & 4.2$\pm$0.2 & 261 &  Cas A\\ 
4.717  & 5.437  & 40241  & 7.8$\pm$0.2 & 10088 & 3.6$\pm$0.1 & 262 & Cas A\\ 
5.631  & 6.850  & 63895  & 6.9$\pm$0.2 & 23911 & 3.2$\pm$0.1 & 262 & ToO\\ 
%7.272  & 7.380  & 6278   & 5.6$\pm$0.7 & 1512  & 2.9$\pm$0.3 & 263 & GPS\\ 
7.689  & 7.801  & 6403   & 4.4$\pm$0.5 & 5675  & 2.7$\pm$0.2 & 263 & Cas A\\ 
7.813  & 9.815  & 117980 & 4.2$\pm$0.1 & 58679 & 2.3$\pm$0.1 & 263 & ToO \\ 
%10.264 & 10.373 & 6219   & 2.6$\pm$0.6 & N/A   & N/A         & 264 & GPS\\ 
10.710 & 12.835 & 106402 & 1.1$\pm$0.1 & 70959 & 0.7$\pm$0.1 & 264 & Cas A\\ 
%13.255 & 13.336 & 4703   & 0.4$\pm$0.7 & 1512  & 0.2$\pm$0.2 & 265 & GPS\\ 
13.700 & 15.575 & 94077  & 0.2$\pm$0.1 & 22281 & 0.1$\pm$0.1 & 265 & Cas A\\ 
\hline 
\noalign{\smallskip} 
\end{tabular} 
\end{center} 
\end{table*} 
 
We use  data from the coded mask imager IBIS/ISGRI \citep{u03,lebr03} 
at energies from 20 to $\sim$ 200 keV (total exposure of 437.4 ks) 
and from the JEM-X monitor \citep{lund03} 
at energies 5 to $\sim$ 20 keV (total exposure 
time of 194.6 ks). 
For ISGRI, the data were extracted for all pointings with a 
source position offset $\leq$ $12^{\circ}$, and for JEM-X with an 
offset $\leq$ $3.5^{\circ}$. 
The spectrometer (SPI) was not used since the source was too weak for
the SPI and its $2^{\circ}$ angular 
resolution is inadequate in this case (see section \ref{sec:analysis}). 
Data reduction was performed using the standard Offline Science 
Analysis (OSA) version 4.2 distributed by the {\it INTEGRAL} Science 
Data Center \citep{c03}. The algorithms used for the spatial and 
spectral analysis are described in \citet{gold03}.

\subsection{RXTE} 
\label{sec:rxte} 
 
After the initial follow-up observations by {\em RXTE} starting 
December 3, 2004, a ToO observation was performed between December 7 
and 21 (observation id. 90425), for which the data were made 
publicly available. For our timing analysis, we also used these data,
namely from the two  
non-imaging  X-ray instruments, the Proportional Counter Array 
(PCA; 2-60 keV) \citep{jahoda96} and the High Energy X-ray Timing 
Experiment (HEXTE; 15--250 keV) \citep{rothschild98}. 
 
The PCA data were collected in the {\tt E\_125us\_64M\_0\_1s} event 
mode, recording event arrival times with 125 $\mu$s time resolution, 
and sorting events in 64 PHA channels. Default selection criteria 
were applied and  the final net PCU2 exposure was 105.5 ks, for 
the  December 7--10 data. We included in our analysis HEXTE ON-source 
data, using defaul screening criteria for Cluster 0 and 1. The 
screened dead-time corrected HEXTE Cluster 0 and 1 exposures were 35.7 
ks and 36.4 ks, respectively, for the December 7--10  data.

\section{Results} 
\label{sec:res} 
 
\subsection{IBIS/ISGRI image and light curves} 
 
%\subsubsection{IBIS/ISGRI} 

\begin{figure} 
%\centerline{\epsfig{file=ima1.ps,width=8cm}} 
\centerline{\epsfig{file=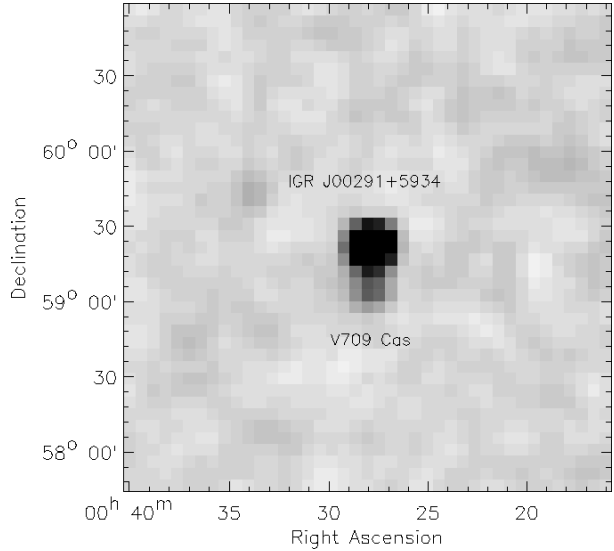,width=8cm}} 
\caption 
{ 
The 20--40 keV ISGRI mosaicked and deconvolved sky image of the 
$\sim437$ ks observation. 
Image size is $\sim 6\fdg25 \times 3\degr$, centered at IGR~J00192+5934 
position. The pixel size is 5$'$. IGR~J00291+5934 and V709~Cas were 
detected at a significance of  $\sim88\sigma$ and $\sim18\sigma$, 
respectively. 
} 
\label{fig:ibis_img} 
\end{figure} 
 
Fig. \ref{fig:ibis_img} shows a significance map in the 20--40 keV 
energy range centered on IGR~J00192+5934 located in the Cas A/Tycho region. 
Single pointings were  deconvolved and analyzed separately, and then 
combined in mosaic images. 
Two sources are clearly detected at a significance level of 
$88.4\sigma$ for \thesource\ and $18.2\sigma$ for the nearby 
intermediate polar V709~Cas. 
In the energy band 40--80 keV, the significance level was 
$51.2\sigma$ for \thesource\ and $6\sigma$ for V709~Cas. At the higher 
energies 80--200 keV the confidence level dropped to 17.1$\sigma$ for 
IGR~J00291+5934,  while V709~Cas was not detected at a statistically 
significant level neither in single exposures nor in the combined 
image. To obtain precise source locations we simultaneously fitted 
the  ISGRI point spread function to the two close sources. We obtained 
a position for \thesource\ at $\alpha_{\rm J2000} = 00^{\rm h}29^{\rm 
m}02\fs92$ and $\delta_{\rm J2000} = 59{\degr}34\arcmin06\farcs4$. The 
position of V709~Cas is given by  $\alpha_{\rm J2000} = 00^{\rm 
h}28^{\rm m}55\fs29$ and $\delta_{\rm J2000}  = 
59{\degr}16\arcmin14\farcs0$. The source position offsets with respect 
to the optical catalog positions \citep{fg04,dws97} are $0\farcm2$ for 
\thesource\ and $1\farcm5$ for V709~Cas. The errors are $0\farcm2$ and 
$1\farcm5$ for \thesource\ and V709~Cas, respectively. These are  within 
the 90\%  confidence level assuming the source location error given by 
\citet{gros03}. The derived angular distance between the two sources 
is $\sim18'$.  Due to the fact that {\em INTEGRAL} is able to image the 
sky at high  angular resolution ($12'$ for ISGRI and $3'$ for 
JEM-X), we were able to clearly distinguish and isolate the 
high-energy fluxes from the two sources separately.  This allowed us to study 
 the X-ray emission of \thesource\ during its entire outburst.

\begin{figure} 
\centerline{\epsfig{file=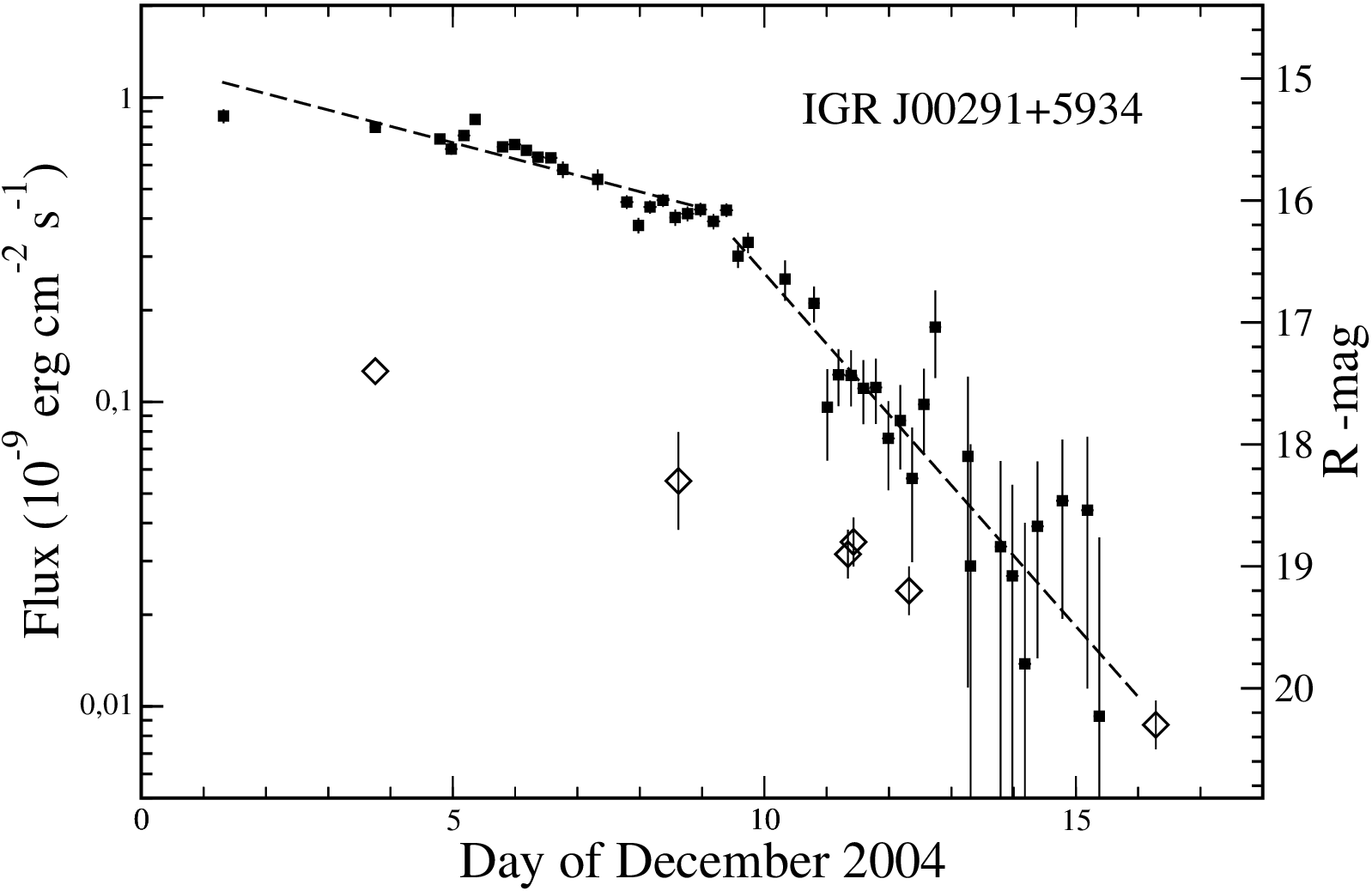,width=9.0cm}} 
%\centerline{\epsfig{file=lc_exp_log.ps,width=6.2cm,angle=-90}} 
\caption 
{ {\em INTEGRAL}/ISGRI light curve in the 20--100 keV energy band 
(averaged over 0.1 day intervals). 
The {\em INTEGRAL}  data  from the whole observation (Table 
\ref{tab:obs})  have been converted to flux assuming a Comptonization model 
(see Table \ref{tab:spec}). The optical flux decay in R-mag is shown 
with diamonds,  where the first two points are taken from 
\citet{fg04} and \citet{sbb04}, respectively, and the rest was taken from 
\citet{bsg04}. The dashed lines correspond to $F \propto 
e^{-t/6.6^{\rm d}}$ and $F \propto e^{-t/2.2^{\rm d}}$. 
} 
\label{fig:decay} 
\end{figure}

The {\em INTEGRAL} 20--100 keV high energy light curve 
has been extracted from the images using all available pointings 
and is shown in Fig.  \ref{fig:decay}, averaged over 0.1 day intervals. 
The first point corresponds to the detection of the source 
\citep{e04}. Using a Comptonization model (see section \ref{sec:spectrum}), 
we estimated the peak bolometric X-ray flux to be $2.1\times10^{-9}$ 
erg cm$^{-2}$ s$^{-1}$. For a distance of 5 kpc (see section 
\ref{sec:companion}), this corresponds to a bolometric 
luminosity of $6.3\times10^{36}$ erg s$^{-1}$, or 3.5\% of the 
Eddington luminosity, $ L_{\rm Edd}$, for a 1.4 M$_{\odot}$ neutron 
star. This is a lower value then measured  for XTE J1751--305 which has a 
luminosity of $\sim 0.13 L_{\rm Edd}$  \citep{gp05},  but similar to 
that of the other accreting X-ray MSPs. 
 
The outburst profile  of \thesource\ (Fig. \ref{fig:decay}) shows 
a decay similar to those of four other millisecond pulsars 
\citep*[e.g.,][]{g98,gp05}. Only 
XTE J1807--294 shows a purely exponential decay with a time scale of 
$\sim$120 days \citep{mf05}. After the 
peak the flux declines exponentially, with a decay time-scale of 6.6 days (10 
days in SAX J1808.4--3658, 7.2 days for XTE J1751-305), until it 
reaches a break, after which the flux drops suddenly with a decay 
time-scale of 2.2 days (1.3 days in SAX J1808.4--3658, $\sim0.6$ days for XTE 
J1751-305).  The higher energy (ISGRI) decay time scale is 
fully consistent with the JEM-X time scale at lower energy (3--20 keV). 
The decay of the optical flux, also shown in Fig.~\ref{fig:decay}, 
is similar to that of the X-ray light curve, but somewhat smoother, with a 
decay time-scale of 4.4 days. Note that the optical source was visible 
up to 45 days after the discovery. 
The ISGRI light curve was also extracted in the following energy ranges: 
20--40 keV, 40--80 keV and 80--200 keV. 
The hardness ratio of these energy bands as a function of time 
indicates that no significant spectral variability was detected.

\subsection{Spectral analysis} 
\label{sec:analysis} 
 
We verified that 
during the whole observation V709~Cas was observed up to 80 keV with a 
constant  mean count rate of $\sim0.6$ in the 20--80 keV energy band, 
and does not influence our spectral analysis for \thesource. 
We performed the spectral analysis using XSPEC version 11.3 \citep{a96}, 
combining the 20--200 keV ISGRI data with the 
simultaneous 5--20 keV JEM-X data. 
A constant factor was included in the fit to take into account the 
uncertainty in the cross-calibration of the instruments. 
A systematic error of 2\% was applied to the JEM-X/ISGRI spectra which 
corresponds to the current uncertainty in the response matrix. 
All spectral uncertainties in the results are given at a 90\% 
confidence level for a single parameter ($\Delta\chi^2=2.71$). 
 
\subsubsection{Total spectrum} 
\label{sec:spectrum} 
 
The joint JEM-X/ISGRI (5--200 keV) 
spectrum  was first fitted with a simple model consisting 
of a photoelectrically-absorbed power-law ({\pl}). Given that we were 
not able to constrain the {\nh} value (as the JEM-X bandpass starts 
above 5 keV) we fixed it to the value found from {\em Chandra} observations 
at lower energies \citep{npw04}. 
A simple {\pl} model is found to be inadequate with {\chiq}/dof=105/36. 
The addition of a cut-off  significantly improves the fit to {\chiq}/dof=47/35, with the best-fit photon index of $\sim$ 1.8 (similar to that reported by 
  \citealt*{shaw05}) 
and a cut-off energy of $\sim$ 130 keV.  However, this model does not describe 
the spectrum  well below 10 keV, which is more complex. 
 
The emission from the accretion shock on a neutron star 
is expected to be produced by thermal Comptonization 
of soft seed photons from the star 
\citep[e.g.][]{zs69,alme73,ls82}. 
Indeed, spectra from accretion-powered MSPs are well described by 
this model \citep{gdb02,pg03,gp05,mf05}. 
 
\begin{figure} 
\centerline{\epsfig{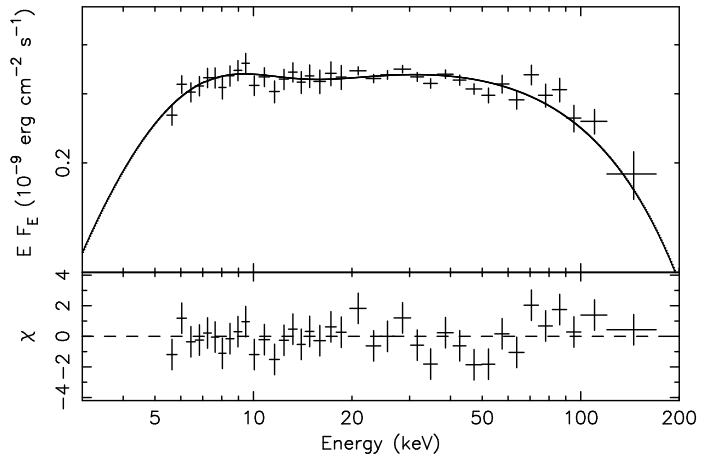}} 
\caption 
{The unfolded spectrum of \thesource\ fitted with an absorbed  {\sc compps} 
model. The data points correspond to the JEM-X (5--20 keV)  and ISGRI 
(20--200 keV) spectra, respectively. The total spectrum  of the model 
is shown by a solid curve. The lower panel presents the residuals 
between the data and the model. 
} 
\label{fig:spec1} 
\end{figure}

We model the shock emission by the {\sc compps} model\footnote{{\compps} 
  is available on\\ 
  ftp://ftp.astro.su.se/pub/juri/XSPEC/COMPPS} \citep{ps96}, 
where the exact numerical solution of the Comptonization 
problem  in different geometries is obtained by 
considering successive scattering orders. 
We approximate the accretion shock geometry by a plane-parallel slab 
at the neutron star surface. The main model parameters are the Thomson optical depth 
  $\tau_{\rm T}$ across the slab, the electron temperature $T_{\rm e}$, 
  and the soft seed photon temperature $T_{\rm seed}$. 
The emitted spectrum depends also on 
the angle between the normal and the line of sight $\theta$ which 
does not coincide with the inclination of the system because of 
light bending. 
The seed photons are injected from the bottom of the slab. 
A fraction $\exp(-\tau_{\rm T}/\cos \theta)$ 
of these photons reaches the observer directly, 
while the remaining part is scattered in the hot gas. 
Thus, the total spectrum contains unscattered black body photons and 
a hard Comptonized  tail. 
The best fit with {\chiq}/dof=44/37 is then obtained  for 
 $kT_{\rm seed}\approx$ 1.5 keV, 
 $kT_{\rm e} \approx 50$ keV,  $\tau_{\rm T} \approx 1.1$, and 
 $\cos \theta \approx 0.6$ (see Table \ref{tab:spec}). 
The model also allows us to determine the apparent area of the seed 
  photons, which turns out to be 
$A_{\rm seed}\sim 21 (D/5\ {\rm kpc})^{2}$ km$^2$ in  the best fit. This 
corresponds to a  hot spot radius of $\sim2.5$ km in this outburst phase. 
The unabsorbed $EF_E$ spectrum and the residuals 
of the data to the model are shown in Fig. \ref{fig:spec1}. 
 
Soft emission (presumable from the accretion disk) with peak 
temperature of about 0.5 keV has been found in other accreting X-ray MSPs 
\citep[see e.g.][]{gp05,mf05} using {\em XMM-Newton} data in the 0.5--10 
keV energy band. As the JEM-X bandpass begins at 5 keV, it is impossible 
to search for this emission in these data. 
Neither the 6.4 keV iron line nor Compton 
reflection were significantly detected.

\begin{table}[htb] 
\begin{center} 
\caption{\label{tab:spec} 
Spectral parameters of the thermal Comptonization 
{\sc compps} fit to the JEM-X/ISGRI data} 
\begin{tabular}{ll} 
\hline\noalign{\smallskip} 
\noalign{\smallskip} 
%Dataset           & JEM-X/ISGRI\\ 
%Model             & {\compps} \\ 
%\hline 
$N_{\rm H} (10^{22} {\rm cm}^{-2})$    & $0.28$ (f) \\ 
%$kT_{\rm disk}$ (keV)    & 0.46$^{+0.11}_{-0.13}$\\ 
%$R_{\rm in} \sqrt{\cos i}$ (km)        & $195^{+142}_{-7}$ \\ 
$kT_{\rm e}$ (keV)       & $49^{+2}_{-6}$\\ 
$kT_{\rm seed}$ (keV)    & $1.49^{+0.16}_{-0.32}$\\ 
$\tau_{\rm T}$           & $1.12^{+0.04}_{-0.07}$ \\ 
$A_{\rm seed}^{a}$ (km$^2$)  & $20.7^{+12.6}_{-4.5}$ \\ 
cos $\theta $            & $0.60^{+0.06}_{-0.09}$ \\ 
$\chi^{2}/{\rm dof}$                   & $44/37$ \\ 
$L_{\rm 1-300 keV}^{a}$  ($10^{36}$ erg s$^{-1}$) & $3.7$\\ 
\noalign{\smallskip} 
\hline 
\noalign{\smallskip} 
\multicolumn{2}{l}{$^{a}$  Assuming a distance of 5 kpc.}\\ 
\end{tabular} 
\end{center} 
\end{table}

\subsubsection{Outburst spectrum} 
\label{sec:outburstspec} 
 
\begin{figure} 
\centerline{\epsfig{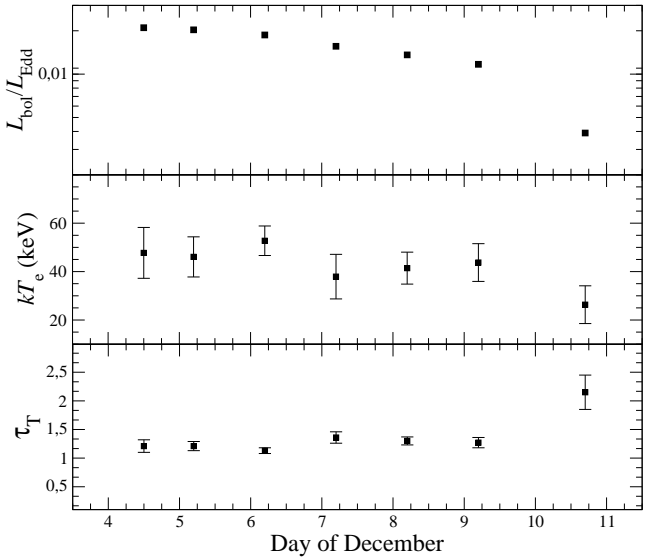}} 
\caption 
{ The outburst evolution of the best-fit spectral parameters of the {\compps} 
  model. Each point, except the last, corresponds to a one day averaged 
  spectrum; due to the lower flux  level the last points are 
  averaged over two days. 
} 
\label{fig:spec2} 
\end{figure}

We analysed the one day averaged JEM-X/ISGRI spectra for the 
observation during the outburst (see Table \ref{tab:obs}). We plot the 
best fit results using the thermal Comptonization {\compps} model in Figure 
\ref{fig:spec2}. 
%The column absorption was fixed at $0.28\times10^{22} {\rm cm}^{-2}$. 
The inclination angle, $\theta$, and the seed photon temperature, 
  $T_{\rm seed}$, were fixed at the best found fit value for the entire 
  observation (see Table \ref{tab:spec}). The luminosity $L_{\rm bol}$ 
 was calculated for a distance 
  of 5 kpc from the best fit model in 
  the energy range 1--300 keV.  
 
The results show that the decay of the outburst is 
marked by a nearly constant plasma temperature, except for the last 
day of outburst where it shows a   decrease along with  an 
increase of the scattering optical depth. Neither variation, however, is 
very  significant statistically.

\subsubsection{Total fluence} 
 
From the broadband spectral model we 
estimated a total fluence from  December 2--15  in the 0.1--300 
keV energy band to be $1.37\times10^{-3}$ erg 
cm$^{-2}$. The fluence estimated with {\em RXTE} is higher by $\sim 
0.4\times10^{-3}$ erg 
cm$^{-2}$  \citep{gmm05}. Owing to the $1^{\circ}$ field of view of 
{\em RXTE}, the flux difference could be due to the intermediate polar V709~Cas, 
located $20'$ from \thesource. The {\it RXTE} pointings were 
stable around $\alpha_{\rm J2000} = 00^{\rm h}29^{\rm m}09\fs23$ and 
$\delta_{\rm J2000} = 59{\degr}33\arcmin58\farcs1$, 
i.e. at every pointing where  both sources were in  the {\it RXTE} 
field of view.  We analysed the spectrum 
of V709~Cas with the identical JEM-X and ISGRI data and exposure times 
reported in Table \ref{tab:obs}. The source was detected from 
5 to 80 keV. In order 
to  estimate the bolometric flux during the outburst period, we fitted 
the spectrum of V709~Cas up to 80 keV with a thermal bremsstrahlung 
model. The best fit with plasma temperature  $\sim$ 20 
keV gives a {\chiq}/dof = 12/15. 
The flux in the energy band 0.1--100 keV was found to be 
$2.1\times10^{-10}$ erg cm$^{-2}$ s$^{-1}$. The flux of V709~Cas was 
constant throughout  the \thesource\ 
 outburst, so the total fluence contributed by V709~Cas from December 2 to 
December 16 was $\sim 0.3\times10^{-3}$ erg cm$^{-2}$. The difference 
between our measured total fluence and that of {\em RXTE} can therefore be 
attributed in large part to the flux of V709~Cas. The remainder is 
likely due to differences in the models used to estimate the source 
flux and uncertainties in the  cross-calibration of the instruments.

\subsection{Timing characteristics} 
 
\subsubsection{The pulsar ephemeris} 
\label{sec:ephemeris}

Searching the low-statistics ISGRI ($E>20$~keV) frequency space for the 
1.67 ms signal from \thesource, involves making numerous trials, which 
reduce substantially the sensitivity to periodic signals. Therefore, we have 
used the high-statistics PCA data (2--60 keV), in order to obtain an 
accurate ephemeris  for the millisecond pulsar. 
 
We first determined the orbital parameters by deriving the instantaneous 
pulse frequency during sufficiently short exposures (as compared to
the orbital  
period) of typically 200 s, in which the pulse profile could 
still be significantly recognized. We applied the $Z_1^2$-statistic 
\citep{b83} to pulse phase distributions for trial frequencies in a 
small window centered on the expected pulse frequency. The PCA data 
until 2004 December 11 (included) yielded sufficiently high 
signals for $\sim 200$ s integration times. The derived distribution 
of frequency and 
frequency uncertainty versus barycentered time (mid of integration 
interval) was subsequently subjected to an epoch folding method 
yielding the ``best", optimized orbital period $P_{\hbox{\scriptsize 
    orb}}$. Folding with this best period resulted in an orbital phase versus 
frequency distribution, which is highly sinusoidal without any significant 
asymmetry. The amplitude resulting from a sine/cosine fit gave the
value of the projected semi-major axis of the  
neutron star orbit $a_{x} \sin i$, and the 
minimum of the fit yielded the time of the ascending node 
($T_{\hbox{\scriptsize asc}}$).

The derived orbital characteristics based on the PCA 
data from December  7--11  are fully consistent with those reported 
by \citet{gmm05} based 
on December 3--6 data. Therefore, in the further analysis we will 
use the set of orbital parameters derived by \citet{gmm05} in order to 
facilitate direct comparisons (note $T_{\rm asc} = 
T_{\pi/2} - 0.25\cdot P_{\rm orb} = 53345.1619261$ MJD TDB). 
 
At this point, we corrected the barycentered time 
tags of the PCA events  for acceleration effects along 
the orbit. For 
each PCA sub-observation  we determined  the 
time of arrival (TOA) by a pulse profile template correlation analysis 
similar to the technique normally used in radio-pulsar studies (see 
e.g. {\tt Tempo},\footnote{http://pulsar.princeton.edu/tempo\/} 
  \citealt*{tw89}). The 
template arrival times could be 
determined accurately only for PCA data collected between 
7 and 12 December 2004 (MJD 53346.202--53351.967), when \thesource\ was 
sufficiently strong. Phase folding the template arrival times 
through a timing model with a constant frequency $\nu = 
598.89213064(1)$ Hz, as found by \cite{gmm05}, gave rise to large 
systematic deviations up to 0.15 in phase from the predicted arrival 
times. A timing model with 2 parameters, $\nu$ and $\dot\nu$, removed 
these deviations (all the residuals, TOAs minus 
model, were $< 0.025$ in phase). 
The fitted timing model has: $\nu = 598.89213060(1)$ Hz; $\dot\nu = 
+8.4(6)\times 10^{-13}$ Hz s$^{-1}$ at epoch MJD 53346 (TDB).
Note that we measure a significant spin-up during this outburst. An
 error in the assumed source coordinates can give rise 
to timing errors which may introduce a spurious spin-up or
spin-down \citep{mt77}. The average 0.092 arcsec  source position error
given by \citet{rd04} would introduce a non-existant spin-up rate of
$\dot \nu = 5.8 \times 
10^{-14}$ Hz s$^{-1}$ during our observation.   Since this
 is  smaller than our measured frequency derivative, we conclude
that the source was spinning up during the outburst. An apparent
spin-up  on the order of our measured $\dot\nu$  would require a
fairly large $\sim$ 0.7 arcsec source position error during our observation.
The ephemeris and orbital parameters are given in Table \ref{tab:ephemeris}.

\begin{table}[h] 
\caption{\label{tab:ephemeris} Ephemeris of \thesource} 
\begin{flushleft} 
\begin{tabular}{ll} 
\hline\noalign{\smallskip} 
\noalign{\smallskip} 
Parameter &  Value  \\ 
\hline 
Right Ascension (J2000)$^{\dagger}$  &  00$^{\rm h}$29$^{\rm m}$3\fs0822 \\ 
Declination (J2000)$^{\dagger}$      &  59$^\circ$$34'$18\farcs99        \\ 
Epoch validity start/end (MJD)       &  53341 -- 53353              \\ 
Frequency                            &  598.89213060(1)   Hz            \\ 
Frequency derivative                 &  $+8.4(6)\times 10^{-13}$  Hz s$^{-1}$ \\ 
Epoch of the period (MJD;TDB)        &  53346                      \\ 
Orbital period                       &  8844.092 s                \\ 
$a_{x} \sin i$                       &  64.993 (lt-ms)                      \\ 
Eccentricity                         &  0            \\ 
Longitude of periastron              &  0$^\circ$    \\ 
Time of ascending node (MJD;TDB)     &  53345.1619261  \\ 
\noalign{\smallskip} 
\hline 
\noalign{\smallskip} 
\multicolumn{2}{l}{$^{\dagger}$ The position is based on radio 
  observations with the }\\ 
\multicolumn{2}{l}{VLA by \citet{rd04}.} \\ 
\end{tabular} 
\end{flushleft} 
\end{table}

\subsubsection{{\em INTEGRAL/RXTE} pulse profiles and time lags} 
\label{sec:pulse} 
 
\begin{figure*}[t] 
{\psfig{figure=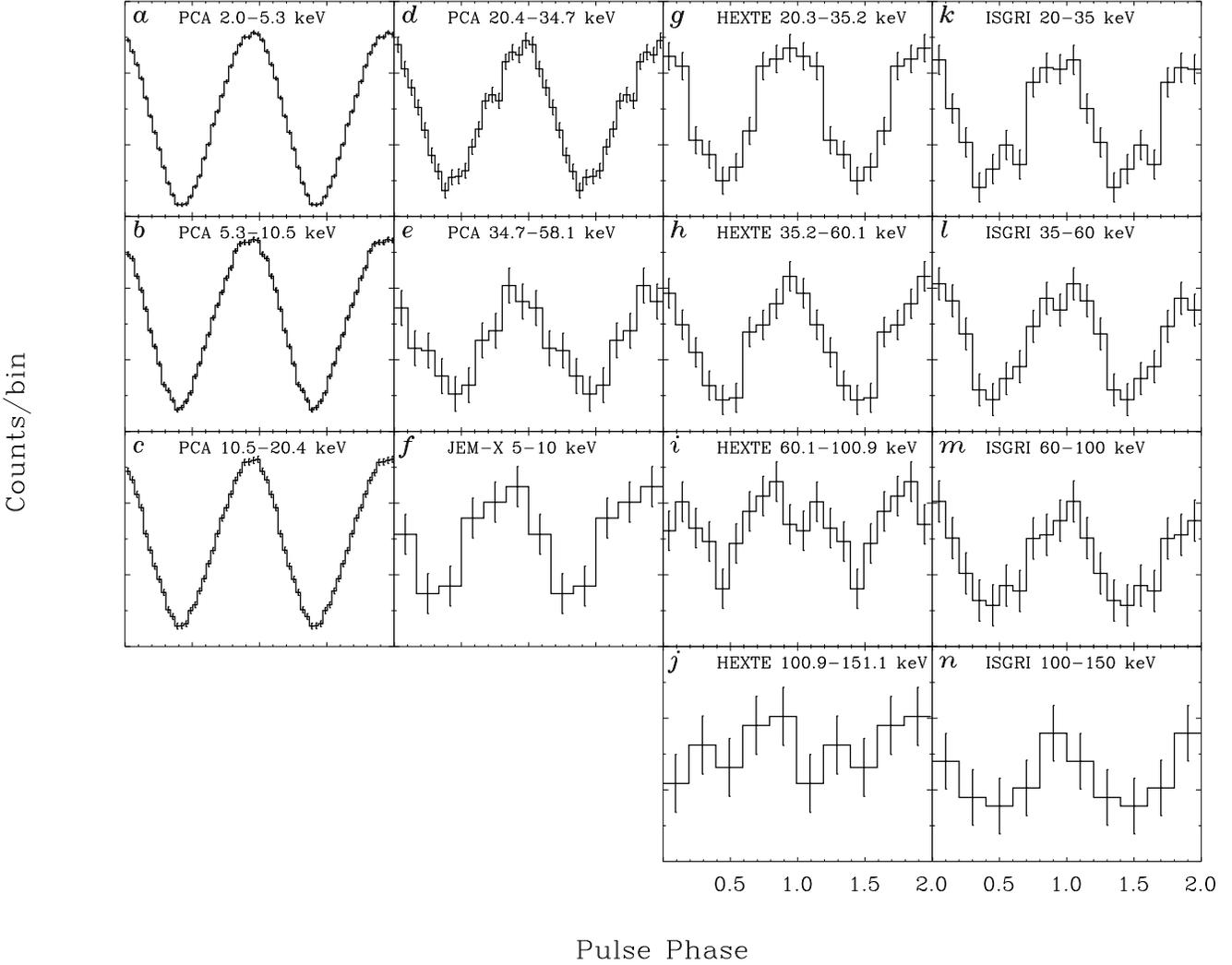,width=15.5cm,angle=90}} 
              {\caption[]{Pulse-profile collage of \thesource\ using data 
              from {\em RXTE}/PCA (2--58 keV; panels a--e), {\em 
              RXTE}/HEXTE (20--151 keV; panels g--j), {\em 
              INTEGRAL}/JEM-X (5--10 keV; panel f) and 
              {\em INTEGRAL}/ISGRI (20--150 keV; panels k--n). Two cycles 
              are shown for clarity. All profiles have high off-set 
              values due to the nature of the instruments. The error 
              bars represent 1 sigma statistical errors. All profiles 
              reach their maximum 
              near phase $\sim 0.95$. Notice the highly sinusoidal 
              shape of the profiles for energies up to 100 keV. 
              \label{fig:profcol}}} 
 
\end{figure*} 
 
We corrected the time stamps of the PCA events from the 2004  December 7--10 (MJD 
53346.185 -- 53350.015) observations to arrival times at the solar system barycenter using 
the JPL DE200 
solar system ephemeris and the position of \thesource\ given 
in Table \ref{tab:ephemeris}, taking into account the binary nature of the 
system. The barycentered times were  phase-folded 
using the pulsar ephemeris given in Table \ref{tab:ephemeris}. 
This yielded pulse phase distributions in 64 energy channels. Pulsed 
emission was detected  up to $\sim 60$ keV. The 
non-uniformity significance derived by applying a $Z_1^2$-statistic is even 
$5.7\sigma$ for the 34.7--58.1 keV band. The pulse profile shapes 
remain highly sinusoidal over the $\sim 2-60$ keV PCA range (see 
Fig. \ref{fig:profcol} panels a--e). 
 
The ISGRI data from {\em INTEGRAL} revolutions 261 -- 263 (see 
Table \ref{tab:obs}) 
were used in our timing analysis aiming at the detection of the 
1.67-millisecond timing signal. {\em INTEGRAL} 
science windows (scw) showing erratic (ISGRI) count rate variations, 
indicative of effects due to Earth radiation belt passage or solar 
flare activity, were excluded from the analysis. 
Only time stamps of events with rise times between 7 and 90 channels 
\citep{lebr03}, detected in non-noisy ISGRI pixels which have an 
illumination factor of more than 25\% were passed for further analysis. 
The event times of these selected events are barycentered, and the 
subsequent pulse phase folding of the barycentered event times using 
$\nu,\dot\nu$ and the epoch of the period (see Table \ref{tab:ephemeris}) 
yielded pulse phase distributions for different energy bands between 
20 and 300 keV. 
 
Significant pulsed emission was detected up to 100 keV, with some 
indication for a signal also between 100 and 150 keV. The $Z_1^2$ 
significance for deviations from uniformity was: $9.1\sigma, 
7.3\sigma, 5.0\sigma$ and $2.0\sigma$ for the  20--35 , 
35--60, 60--100 and 100--150 keV energy bands, respectively. 
The sinusoidally shaped pulse profiles are shown in the panels k--n of 
Fig.~\ref{fig:profcol}. 
 The 1.67 ms timing signal was also  detected in data from 
the JEM-X detector: $4\sigma$ non-uniformity significance 
for the 5--10 keV band (no event selections except on energy were 
applied; see Fig.~\ref{fig:profcol} panel f). 
 
To complement and verify our ISGRI timing results we also 
produced pulse phase distributions based on HEXTE (15--250 keV) 
data from the same time period as the PCA events above. The HEXTE 
profiles are shown in panels g-j of Fig.~\ref{fig:profcol}. The $Z_1^2$ 
significances were: $11.4\sigma, 8.3\sigma, 3.3\sigma$ and $1.1\sigma$ 
(=no detection) for the energy bands, 20.3--35.2, 35.2--60.1, 60.1--100.9 
and 100.9--151.1 keV, respectively. 
 
Note that the pulse profiles detected by the PCA, 
JEM-X, HEXTE and ISGRI are fully consistent in shape and absolute 
timing. \citet{gmm05} reported an energy-dependent time delay, with 
the 6-9 keV pulses arriving up to $85{\mu}$s earlier than those at 
lower energies (see Fig. 2 bottom panel of Galloway et al. 2005). 
Exploiting the ISGRI and HEXTE pulse phase 
distributions above 20 keV, we can investigate this behavior up to $\sim 100$ 
keV. We performed a similar analysis, first determining the 
time lags between profiles in different PCA energy bands, but now for 
the independent PCA dataset from 2004 December 7--10. We confirm the 
reported variation in time  lag between $\sim 2 - 20$ keV as a function 
of energy \citep{gmm05} and  extend the PCA coverage up to $\sim 35$ keV. 
 
\begin{figure}[t] % 
              {\hspace{-0.3cm}\psfig{figure=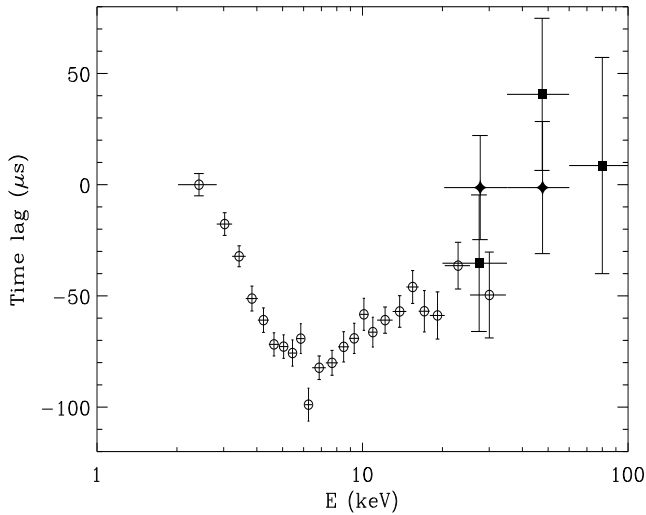,width=9.cm,height=7.5cm}} 
              {\caption[]{Time lags as a function of energy in the 2--100 keV 
               energy range combining {\em RXTE}/PCA (2--35 keV; 2004 
              December 7--10; open circles), {\em RXTE}/HEXTE (20--60 keV; 
              2004  December 7--10; filled diamonds) and {\em 
              INTEGRAL}/ISGRI (20--100 keV; December  3.6--9.8; filled 
              squares) measurements. 
                \label{fig:timelag}} 
              } 
\end{figure}

Correlating the three significant  ISGRI and two significant 
HEXTE pulse profiles between 20 keV and 100 keV with the low energy 
(reference) PCA profile (2.02-2.82 keV), yielded the time lags as 
measured with ISGRI and HEXTE above 20 keV.  Note, that a time shift 
of $+0.5\times 125\mu$s has been applied to the PCA time stamps, 
because the PCA times refer to the start of the time pixel instead of 
the mid in the case of the {\tt E\_125us\_64M\_0\_1s} event mode. Both 
the PCA/HEXTE and  ISGRI time lags as a function of energy 
are shown in Fig. \ref{fig:timelag}, covering the 2--100 keV energy band. 
We see that the ISGRI and  PCA/HEXTE time lag measurements in 
the overlapping 20--35 keV energy band are fully consistent, and that 
the time lags  decrease for increasing energies, attaining values 
consistent with zero for energies in excess 
of 35 keV consistent with a trend already set in near $\sim 6$ keV. 
The full consistency of the PCA/HEXTE and ISGRI time lags in 
the overlapping 20--35 keV band indicates that the time alignment 
between {\em RXTE} and {\em INTEGRAL} is better than $\sim 25\mu$s. 
 
\subsubsection{Spectrum of the pulsed emission and pulsed fraction} 
\label{sec:specpulsed} 
 
Important diagnostic parameters for constraining the parameter space 
in theoretical modelling \citep*[see e.g.][]{vp04} are the pulsed 
spectrum and the pulsed fraction (here defined as: pulsed flux/total 
flux) as a function of energy. While the pulsed fluxes can be derived 
for all the main instruments considered in this study, the total 
 flux of \thesource\ can  only be derived reliably for 
the high-energy instruments with arcminute imaging 
capabilities. In particular, the nearby source V709~Cas 
contaminates the total flux estimate of \thesource\ using the non-imaging 
{\em RXTE} instruments (see section \ref{sec:spectrum}). Galloway et 
al. (2005) did 
use the PCA to derive the fractional rms amplitude of the 
pulsations as a function of 
energy (in their Fig. 2, top panel); therefore, their values contain a 
contribution from V709~Cas and cannot be used to estimates the fractional 
rms amplitude of \thesource.

\begin{figure}[t] 
%{\hspace{-0.25cm}\psfig{figure=idl_IGRJ00291+5934_HE_SPC_UPDATE_KEV.ps,width=9.cm,height=8cm}} 
{\hspace{-0.25cm}\psfig{figure=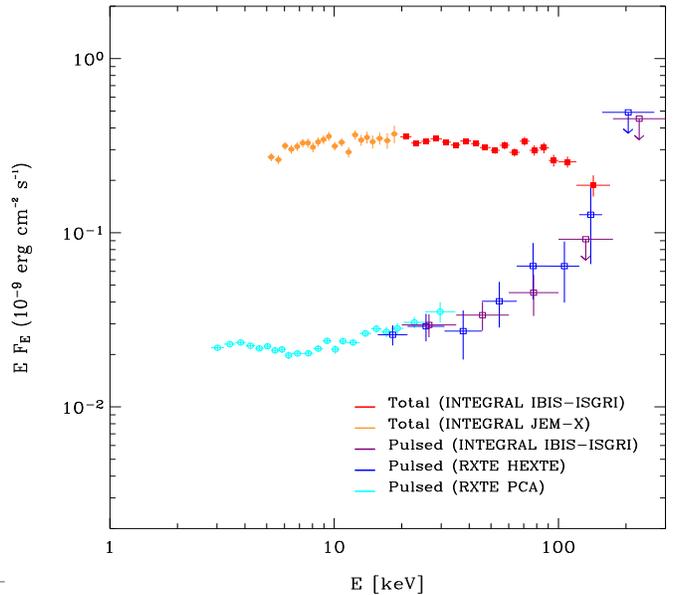,width=9.cm,height=8cm}} 
              {\caption[]{The total and pulsed spectrum of \thesource\ in an 
              $EF_{\rm E}$ representation from 3 to 300 keV combining 
              flux measurements from {\em RXTE}/PCA/HEXTE and {\em 
              INTEGRAL}/ISGRI for the pulsed part and {\em INTEGRAL}/ISGRI and 
              JEM-X for the total part. Notice the hardening of the 
              pulsed spectrum towards higher energies. 
                 \label{fig:pulspectrum}} 
              } 
\end{figure}

We derived the pulsed fluxes for the 
PCA, HEXTE and ISGRI instruments (the JEM-X statistics for the 
pulsed component is too low, cf. Fig.~\ref{fig:profcol} panel f) and the 
total fluxes only for the imaging {\em INTEGRAL} instruments\footnote{The 
  {\em RXTE} and {\em INTEGRAL} measurements were nearly contemporaneous.}. 
To derive the pulsed fluxes we determine the number of pulsed excess 
counts in a given energy band.  A truncated 
Fourier-series fit using 3 harmonics to the pulsed phase distributions 
as a function of energy yielded this quantity. In case of the PCA the 
pulsed excess counts (2--30 keV) have been converted to flux values 
(ph/cm$^2$s\,keV) in a forward energy spectral folding method assuming 
an underlying absorbed ($N_{\rm H}=2.8\times 10^{21}$ cm$^{-2}$; 
see \citealt*{npw04}) \pl\ model with an energy dependent index taking 
into account the different exposures and energy response of the 5 
different PCUs. Figure \ref{fig:pulspectrum} shows these PCA pulsed flux 
measurements in an $EF_{\rm E}$ representation.

\begin{figure}[t] 
%              {\hspace{-0.25cm}\psfig{figure=idl_IGRJ00291+5934_PULSED_FRACTION.ps,width=9.5cm,height=7.5cm}} 
 {\hspace{-0.25cm}\psfig{figure=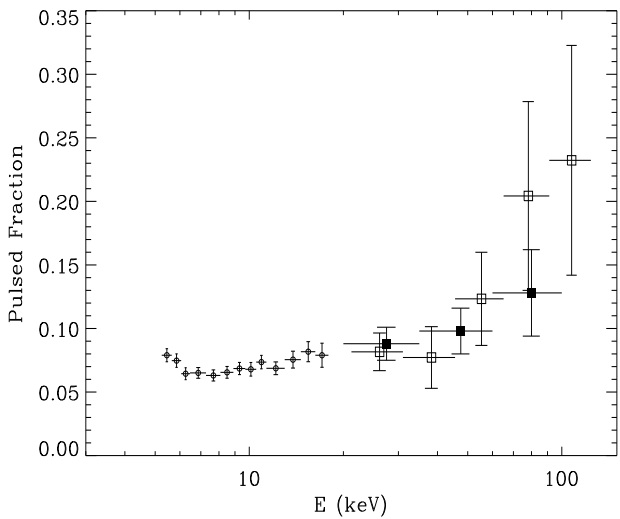,width=9.cm,height=7.5cm}} 
              {\caption[]{The pulsed fraction (=pulsed flux/total 
              flux) of \thesource\ using pulsed/total flux measurements from 
               PCA/HEXTE and  JEM-X/ISGRI (open 
              circles, PCA pulsed flux relative to JEM-X total flux; 
              open squares, HEXTE pulsed flux relative to ISGRI 
              total flux; filled squares, ISGRI pulsed flux 
              relative to ISGRI total flux). The pulsed fraction 
              gradually increases from $\sim 6\%$ at $\sim 6$ keV to 
              $\sim 12-20\%$ at $\sim 100$ keV. 
                 \label{fig:pulsefraction}} 
              } 
\end{figure}

For HEXTE an equivalent method -- dividing the pulsed excess count 
rates in a certain energy band by its effective sensitive area 
assuming a \pl\ model -- was used to obtain the pulsed fluxes in 
the 15.6--267.5 keV energy range taking into account the different 
cluster 0 and 1 dead time corrected ON exposures and energy responses 
(see Fig. \ref{fig:pulspectrum}). 
 
Finally, for ISGRI the pulsed excess counts (20--300 keV), 
accumulated during an exposure of 336.94 ks, have been converted to 
flux values using (a) the pulsed excess counts of the Crab pulsar as a 
reference from an equivalent $5\times5$ on-axis dither observation 
(Rev-102; 54.33 ks), determined in the same energy bands as chosen for 
\thesource, and (b) a spectral model for the pulsed emission of the Crab 
pulsar \citep*[see e.g.][]{k01} over the 15--10\,000 keV energy 
range based on HEXTE and {\em CGRO}/COMPTEL measurements. These ISGRI 
pulsed flux measurements are also shown in Fig.~\ref{fig:pulspectrum}.

From the PCA/HEXTE and ISGRI pulsed flux measurements shown in 
Fig. \ref{fig:pulspectrum} it is clear that this emission component is 
hardening towards higher energies. Superposing now also the total 
\thesource, spectrum from JEM-X and ISGRI 
measurements (see section \ref{sec:spectrum}) in Fig. \ref{fig:pulspectrum}, 
which softens weakly towards higher energies, it is clear that the 
pulsed fraction 
increases as a function of energy. The energy dependence of the pulsed 
fraction is shown in Fig.~\ref{fig:pulsefraction}. It gradually increases 
from $\sim 6\%$ at 6 keV to $\sim 12-20 \%$ near 100 keV. It is the 
first time that such behaviour has been measured from any of the known 
accretion-powered ms pulsars. 
 
We  note here that \citet{gmm05} report decrease of the fractional 
rms amplitude  of oscillations  with increasing energy above 10 keV. 
As remerked befor, this can be explained by the contribution in the PCA data 
from V709~Cas, which has a harder spectrum than \thesource\ in this 
energy range. 
 
%\citet{gmm05} also find decreasing of the rms amplitude during the outburst. 
%We have studied the rms behaviour during the outburst with the ISGRI data 
%and find 
 
\section{Discussion} 
 
\subsection{Companion star and the distance} 
\label{sec:companion} 
 
Taking the {\em INTEGRAL} observed total fluence of 
$1.37\times10^{-3}$ erg cm$^{-2}$, and a recurrence time $T$ of 3 
years \citep{rr04}, we find a long term time-averaged flux of 
1.45$\times$10$^{-11}$ erg cm$^{-2}$ s$^{-1}$ (smaller than the 
1.8$\times$10$^{-11}$ erg cm$^{-2}$ s$^{-1}$ flux reported by 
\citealt{gmm05}). Following the current model of \citet{vh95} and \citet{bc01}, we 
therefore find,  for an inclination $i = 90^\circ$ and  minimum 
companion mass $M_{\rm c} = 0.039\ {\rm M}_\odot$ 
(for a 1.4 ${\rm M}_\odot$ neutron star), 
 a lower limit on the source distance of 4.7 kpc. 
At a most probable inclination angle $i = 60^\circ$, we obtain 
a source distance of 5.4 kpc. In both cases the companion  is  a hot 
brown dwarf. The $\sim$1 kpc difference between our calculated minimum 
source distance and that of \citet{gmm05} is attributable to the 
difference in observed total flux.  At a distance of 
$\sim$ 5 kpc, \thesource\ lies about 270 pc off the galactic plane, 
placing it within the thick disk, but outside the thin disk.

\subsection{Origin of X-ray emission} 
 
The spectral parameters obtained from our thermal Comptonization fits 
to the broad-band data for 
 \thesource, $\kte=49$ keV and $\taut=1.12$, 
 are similar to those of other MSPs: 
 $\kte=60$ keV, $\taut=0.88$ in SAX~J1808.4-3658, 
 $\kte=33$ keV, $\taut=1.7$ in XTE J1751-350   \citep[see][]{gp05}. 
We note that the product $\taut\times\kte$ is amazingly close 
in the three sources. 
The spectral shape and the product $\taut\times\kte$ 
are very stable during the outburst as is observed 
in other sources too \citep{g98,gp05}.

The constancy of the spectral slopes  during 
the outbursts and their extreme similarity 
in different MSPs can be used as an argument that the 
emission region geometry does not depend on the accretion rate. 
If the  energy dissipation takes place in a hot shock, while the cooling 
of the electrons (that emit X/$\gamma$-rays via thermal Comptonization) 
is determined by the reprocessing of the hard 
X-ray radiation at the neutron star surface (so called two-phase model, see 
\citealt{hm93,stern,p98,tlm98,mbp01}), 
the spectral slope is determined by the energy balance in the hot phase 
and  is a function of the geometry. At constant geometry (e.g. slab), 
the temperature depends on the optical depth, but $\taut \times \kte$ 
is approximately constant.

\subsection{Pulsed fraction and pulse profile} 
 
%We also report the dependance of the pulsed fraction on energy using three pulsed 
%spectra (from the PCA, HEXTE and ISGRI), and using the total JEM-X and ISGRI spectra. 
We showed that the pulsed fraction gradually increases with energy 
from $\sim 6\%$ at 6 keV to $\sim 12-20 \%$ near 100 keV. For the 
first time such behaviour has been measured for any of the accretion-powered MSPs. 
 
A spot emitting as a black body (with isotropic specific intensity) 
at a slowly rotating star produces 
nearly sinusoidal variations in the observed flux \citep{belo02}, 
and the peak is reached when the spot is closest to the observer. 
When the spot radiation is anisotropic, 
the  harmonics appear in the pulse profile, 
with their strength  depending on the degree of anisotropy \citep{poutanen04,vp04}. 
In case of rapid rotation, as in MSPs, 
Doppler boosting affects the variability pattern. 
When the spot moves towards the observer, the emission 
increases, while for a spot moving away, the flux drops. 
The Doppler factor reaches the maximum a quarter of the period before the 
peak of the projected area  shifting the emission 
peak towards earlier phase. 
The observed flux due to the Doppler effect varies 
as the Doppler factor in power $(3+\Gamma)$ 
 \citep{pg03,vp04}, where the photon index $\Gamma$ could be a function 
of energy. If the Doppler factor varies around 1 with 2\% amplitude, 
we get 10\% variability. 
 
The Comptonized spectrum can be approximated as 
\be 
F_E \propto E^{-(\Gamma_0-1)} \exp\left( -[E/E_{\rm c}]^{\beta}\right) , 
\ee 
where  $E_{\rm c}\sim \kte$ is the energy of the cutoff  and parameter 
$\beta\sim2$ describes its sharpness. 
The local photon index is then 
\be 
\Gamma(E)\equiv 1 - \frac{\mbox{d} \ln F_E }{\mbox{d} \ln E} = 
\Gamma_0 + \beta (E/E_{\rm c})^{\beta} . 
\ee 
At low energies, $\Gamma\approx\Gamma_0$, and 
rms (or pulsed fraction) is a very weak function of energy. 
Close to the cutoff, the spectral index rapidly increases and 
the pulsed fraction should grows with energy, as observed (see 
Fig.~\ref{fig:pulsefraction}). 
The Comptonization models predict  softening of the total spectrum 
with simultaneous hardening of the pulsed spectrum at higher energies. 
 
The observed pulse profiles are almost purely sinusoidal 
\citep[see Fig.~\ref{fig:profcol} and also][]{gmm05}. 
The total variability amplitude  is proportional to the product 
$\sin i\ \sin \theta$ \citep{poutanen04}, 
where $\theta$ is the angular distance of the spot from rotational axis. 
The relative amplitude of the second harmonics to the fundamental is also 
proportional to the same product. 
%Thus, for a rather small total variability one does not expect strong 
%harmonics either. 
The $\sim 6\%$ observed amplitude constrains the angles  $i$ and $\theta$ 
\citep[see e.g.][]{gp05}. We estimate that for a 1.4$M_\odot$ neutron 
star with 12 km radius, $\theta$ varies between $\sim3^\circ$ and $15^\circ$ 
when the inclination $i$ varies from $90^\circ$  to $18^\circ$.

\subsection{Time lags} 
 
The observed spectrum of \thesource\ consists of a black body 
from the neutron star surface and a component produced by 
Comptonization of these seed photons in the hot electron  region, 
presumably a shock which can be represented as a plane-parallel slab. 
The angular distributions of the black body and Comptonized 
photons emitted by the slab are significantly different. 
The difference in the emission patterns causes the two components 
to show a different variability pattern as a function of the pulsar phase, 
with the hard Comptonized component leading the soft black body 
component \citep{pg03,vp04}. 
 
This scenario is consistent with observations of 
SAX~J1808.4-3658 \citep{cmt98,gdb02}, XTE~J1751-350 \citep{gp05}, 
where time lags increase rapidly with energy until 7-10 keV, 
where the black body contribution becomes negligible 
and then saturate above 10 keV. In \thesource, however, 
time lags first increase until 7 keV and then decrease slightly 
(see Fig.~\ref{fig:timelag} and \citealt*{gmm05}). Judging from the PCA data, 
it seems that the lags saturate above 15 keV. 
The decrease of the time lags above 7 keV does not have an obvious 
explanation. 
There are also (not very significant) 
indications  from the HEXTE and ISGRI data that the lags even 
reach zero value at $\sim50$ keV. If confirmed, this kind of behaviour 
would be a serious challenge to any model.

\section{Summary} 
 
We analyzed the spectral and the timing behaviour from the entire {\em 
 INTEGRAL} observation of \thesource, which was spatially well distinguished 
 from the neighboring source V709~Cas. 
 The time averaged broad-band spectrum from 5 to 300 keV obtained with 
 JEM-X/ISGRI is well described by a 
 thermal Comptonization model with  seed photons from the neutron star 
  surface scattered in a shock-heated accretion column above the hot spot 
  region. Additionally, studying the spectral evolution, we found that the 
  electron temperature and  the related optical depth are almost 
 invariant  as the  flux decreases  during the  outburst. 
 
 The {\em INTEGRAL}/ISGRI data, supported by simultaneously obtained 
  {\em RXTE}/PCA and HEXTE data,  also allowed us 
  to study the timing behaviour in the 
 millisecond range for energies up to 150 keV. 
 We detected the 1.67 ms period consistently with all three instruments 
 and found that the pulse profile is close to a sine function 
 at all  studied energies 
 as observed for the other MSPs \citep[see e.g.,][]{gdb02,gp05}. 
 This is the first 
 time that pulsed emission has been detected from 2 up 
to $\sim 150$ keV for any of the currently known accretion-powered 
millisecond pulsars. Moreover, it demonstrates the excellent timing 
capabilities of {\em INTEGRAL} \citep[see][]{k03} down to the 0.1 ms 
level. 
We have discovered that the pulsed fraction significantly 
increases with energy which is naturally explained by 
the effect of Doppler boosting on the  exponentially cutoff Comptonization 
spectrum. We have measured a  spin-up rate of  +8.4$\times$ 10$^{-13}$ 
  Hz s$^{-1}$. Accurate measurements of the frequency derivative are 
  necessary to constrain models of NS spin evolution in X-ray binaries 
  \citep[e.g.,][]{lw04}. 
 
From our total fluence measurements 
assuming that the mass accretion rate is 
driven by gravitational radiation, 
we find the minimum source distance of 4.7 kpc. 
%for  the most probable inclination angle of $i = 60^\circ$, 
%we find the source distance to be $\sim$5 kpc. 
%indicating a $\sim 0.042 {\rm M}_\odot$ hot brown dwarf companion star. 
This locates the source, like other 
accreting X-ray MSPs, within the thick disk of the Galaxy.

\acknowledgements 
MF acknowledges the CNRS for financial support and  thanks 
A. Decourche and J. Vink for the CAS A/Tycho INTEGRAL A02 data. 
LK \& WH are supported by the Netherlands Organisation for Scientific 
Research  (NWO), and are grateful to A. Rots, J. in 't Zand for valuable 
discussions. JP acknowledges the Academy of Finland grant 201079. 
MF and  JP were supported in part by 
the NORDITA Nordic project on High Energy Astrophysics.

\end{document}